\documentclass[epj,twocolumn]{webofc}
\usepackage[varg]{txfonts}   
\woctitle{}
\begin{document}
\title{Dark Matter 2014}

\author{Marc Schumann\inst{1}\fnsep\thanks{\email{marc.schumann@lhep.unibe.ch}}}
        
\institute{Albert Einstein Center for Fundamental Physics, University of Bern, Switzerland}

\abstract{
  This article gives an overview on the status of experimental searches for dark matter at the end of 2014. The main focus is on direct searches for weakly interacting massive particles (WIMPs) using underground-based low-background detectors, especially on the new  results published in 2014. WIMPs are excellent dark matter candidates, predicted by many theories beyond the standard model of particle physics, and are expected to interact with the target nuclei either via spin-independent (scalar) or spin-dependent (axial-vector) couplings. Non-WIMP dark matter candidates, especially axions and axion-like particles are also briefly discussed.
}
\maketitle
%
\section{Introduction}\label{intro}

Numerous indirect observations at astronomical and cosmological scales~\cite{ref::evidence}, as well as results from complex numerical $N$-body simulations~\cite{ref::simulation}, indicate the presence of a new form of matter in the Universe, which only interacts significantly via gravity. The existence of this \emph{dark matter} is one of the strongest indications for physics beyond the standard model of particle physics, as no known particle can be attributed to dark matter.
Recently, results from the Planck satellite mission~\cite{ref::planck} on precise measurements of the cosmic microwave background (CMB) have been published. These agree with the predictions of the $\Lambda$CDM model, describing a cosmos dominated by dark energy ($\Lambda$) and cold dark matter (CDM). According to Planck, 68.3\% of the Universe's energy density is from dark energy and from 26.8\% dark matter.

The particle(s) constituting the dark matter remain unknown as of today, even though dark matter outnumbers 'ordinary' baryonic matter by a factor~5. Neutral, cold (i.e.~non-relativistic), and stable particles (or with half-lifes longer than the age of the Universe), representing viable dark matter candidates, are predicted by many theories beyond the standard model, especially if their decay is inhibited by a new symmetry. Important examples are the lightest supersymmetric particle (LSP) in in supersymmetric theories~\cite{ref::susy}, such as the neutralino $\chi_0$, the lightest $T$-odd particle in little Higgs theories~\cite{ref::littlehiggs}, or the lightest Kaluza-Klein particle (LKP) in models with extra-dimensions~\cite{ref::extradim}. This class of dark matter candidates are referred to as weakly interacting massive particles (WIMPs)~\cite{ref::wimps}, covering a largely unconstrained mass range from 1\,GeV/$c^2$ to several~TeV/$c^2$.

This article attempts to provide an overview of the experimental status of direct WIMP searches at the end of~2014, focusing on the projects leading the field and on new results published during the year. We do not intend to provide detailed descriptions of the various experiments, but mainly focus on the underlying concepts and the recent results, and refer the reader to the references for further information. This review is a personally biased selection and other authors might place their focus differently. 

A short review on the formalism to describe WIMPs scattering in lab-based detectors and the relevant backgrounds is given in Section~\ref{sec::direct}. Sections~\ref{sec::wimps} and~\ref{sec::other} summarize the status and the main new results on WIMP and important non-WIMP searches, respectively, followed by a brief conclusion.

\section{Principles of Direct Detection}
\label{sec::direct}

If the WIMP dark matter particle interacts not only gravitationally with ordinary matter, but also with weak-scale cross sections, the signature of WIMPs present in the local solar neighborhood might be directly detectable in sensitive Earth-based detectors~\cite{ref::baudis}. This path towards a WIMP observation is called \emph{direct detection}. Alternative approaches are the \emph{indirect detection} of WIMPs by observing their annihilation (or decay) products (eventually leading to $\gamma$, $\nu$, $e^+$, $\overline{p}$, etc.) in satellite experiments and the production of dark matter in \emph{collider experiments}.

In this article, we concentrate on direct detection.

\subsection{Event Rates and Energies}

Neutral WIMPs are expected to scatter off the target nuclei of a direct detection experiment of mass $M$. The  event rate per nuclear recoil (NR) energy $E_r$ is then given by
\begin{equation}\label{eq::rate}
\frac{dR}{dE_r} = \frac{\rho_0 M}{m_N m_\chi} \int_{v_{min}}^{v_{esc}} v f(v) \frac{d \sigma}{dE_r} \ dv\textnormal{.}
\end{equation}
$m_N$ and $m_\chi$ are the masses of target nucleus and WIMP particle, respectively, and $f(v)$ is the normalized WIMP velocity distribution. All velocities are defined in the detector's reference frame, with \begin{equation}\label{eq::mu}
v_{min}=\sqrt{\frac{E_r m_N}{2} \frac{(m_N+m_\chi)^2}{(m_N m_\chi)^2}} = \sqrt{\frac{E_r m_N}{2} \frac{1}{\mu^2}}                                                                                                                                                                                                                        
\end{equation}
being the the minimal velocity required to induce a nuclear recoil $E_r$. The escape velocity $v_{esc}=544$\,km\,s$^{-1}$~\cite{ref::rave} is the maximum velocity for WIMPs bound in the potential well of the galaxy. The canonical value for the local WIMP density used for the interpretation of measurements is $\rho_{0}=0.3$\,GeV/$c^{2}$/cm$^3$. The observed number of events in an experiment running for a live-time $T$ is obtained by integrating Eq.~(\ref{eq::rate}) from the threshold energy $E_{low}$ to the upper boundary  $E_{high}$:
\begin{equation}
N=T \int_{E_{low}}^{E_{high}} dE_r \ \epsilon(E_r) \ \frac{dR}{dE_r} \textnormal{,} 
\end{equation}
with the detector efficiency $\epsilon$. As $dR/dE_r$ is a steeply falling exponential function with $E_r = {\cal O}$(10)\,keVr only, $E_{high}$ is much less relevant than the energy threshold $E_{low}$. The nuclear recoil energy is given in keVr (nuclear recoil equivalent), which is different from the electronic recoil scale (keVee) due to quenching effects caused by the different energy-loss mechanisms.

Because of its large de Broglie wavelength, the WIMP interacts coherently with all nucleons in the target nucleus. The WIMP-nucleus scattering cross section in Eq.~(\ref{eq::rate}) is velocity and recoil-energy dependent and given by
\begin{equation}
\frac{d\sigma}{dE_r} = \frac{m_N}{2 v^2 \mu^2} \left( \sigma_{SI} F_{SI}^2(E_r) + \sigma_{SD} F_{SD}^2(E_r) \right)\textnormal{.}
\end{equation}
The loss of coherence is accounted for by the finite form factors $F_i$, which are only relevant for heavy WIMP targets such as Xe or I. Since the interaction of WIMPs with baryonic matter is a priori unknown, the cross section consists of two terms, for spin-independent (SI, scalar) and spin-dependent (SD, axial-vector) couplings. The first one reads
\begin{equation}\label{eq::si}
\sigma_{SI}=\sigma_n \frac{\mu^2}{\mu_n^2}\frac{(f_pZ+f_n(A-Z))^2}{f_n^2} = \sigma_n \frac{\mu^2}{\mu_n^2} A^2 \textnormal{,}
\end{equation}
where the $f_{p,n}$ describe the WIMP couplings to protons and neutrons, and the second equality assumes $f_p=f_n$, leading to a $A^2$ dependence of the cross section. $\mu$ is the WIMP-nucleus reduced mass, see Eq.~(\ref{eq::mu}), and $\mu_n$ the one of the WIMP-nucleon system. It is used to relate the WIMP-nucleus cross section $\sigma$ to the WIMP-nucleon cross section $\sigma_n$ (which allows for comparisons between different target nuclei).
The differential spin-dependent cross section reads
\begin{equation}\label{eq::sd}
\frac{d\sigma_{SD}}{\textnormal{d}|\vec{q}|^2} = \frac{8 G_F^2}{\pi v^2} \left[ a_p \langle S_p \rangle + a_n \langle S_n \rangle \right]^2 \frac{J+1}{J} \frac{S(|\vec{q}|)}{S(0)} \textnormal{.} 
\end{equation}
The $\langle S_{p,n} \rangle$ are the expectation values of the total spin operators in the the nucleus. The spin-dependent case shows no $A^2$ enhancement of the cross section, but is related to the total nuclear spin $J$ of the target nucleus as well as its spin-structure function $S(|\vec{q}|)$, which depends on the momentum transfer $\vec{q}$. While heavy nuclei are generally more sensitive to SI-interactions (ignoring important detector details such as threshold and background), this is not true for the SD-case. As neutrons and protons in the target can contribute differently to the total spin, one usually quotes the SD-results assuming a WIMP-coupling to protons ($a_n=0$) and neutrons ($a_p=0$) only.

While the shape of the expected recoil spectrum is totally determined by kinematics ($f(v)$, $m_N$, $m_\chi$), the total expected rate depends on the cross section (for a given $\rho_0$). Spin-independent rates are of $<$1\,event/kg/year have already been excluded, corresponding to WIMP-nucleon cross sections $\sigma_n\approx10^{-45}$\,cm$^2$ for $m_\chi \sim 50$\,GeV/$c^2$. In order to be sensitive to such cross sections, the optimal WIMP detector should therefore have a large total mass $M$, a high mass number $A$, a low energy threshold $E_{low}$, an ultra-low background and the ability to distinguish between signal and background events.

\subsection{Backgrounds}

Most current WIMP searches are dominated by $\gamma$-back\-grounds from the environment or the experimental setup itself and by $\beta$-particles at the surfaces or in the bulk of the detector. They generate electronic recoils (ER) by electromagnetic interactions with the atomic electrons. The different ionization density of ERs compared to the WIMP-induced nuclear recoils (NR) is often used to discriminate background (ER) from signal (NR). $\alpha$-contamination in the detector materials is usually uncritical and only becomes relevant if the major part of the $\alpha$-energy is lost in insensitive detector regions. (A notable exception are the bubble chambers described in Section~\ref{sec::sd}.)

The most dangerous background for direct detection experiment are single-scatter neutron-induced nuclear recoils from ($\alpha,n$) and spontaneous fission reactions, or induced by muons, as such events cannot be distinguished from a WIMP signal. The event multiplicity is crucial as WIMPs will only scatter once in the detector due to their tiny cross section, while neutrons have a shorter mean-free path and will often generate multiple-scatter signatures. Effective WIMP detectors can therefore identify (and reject) multi-scatter events. If the interaction vertex can be additionally measured with some precision, the background can be further reduced by exploiting the self-shielding capability of the target material, as background events predominantly occur close to the detector surfaces. This \emph{fiducialization} is especially effective for high-$Z$ materials.

Massive shields around the detector, which itself is made from selected low-background materials, are used to suppress most backgrounds: high-$Z$ materials such as lead and copper or large amounts of water are efficient against external $\gamma$-rays, polyethylene and water against neutrons. In order to reduce muon-induced neutrons, dark matter detectors are installed in deep-underground laboratories: their typical rock overburden of 1-2\,km suppresses the muon flux by 5-7\,orders of magnitude.

\begin{figure*}
\centering
\includegraphics[width=0.8\textwidth]{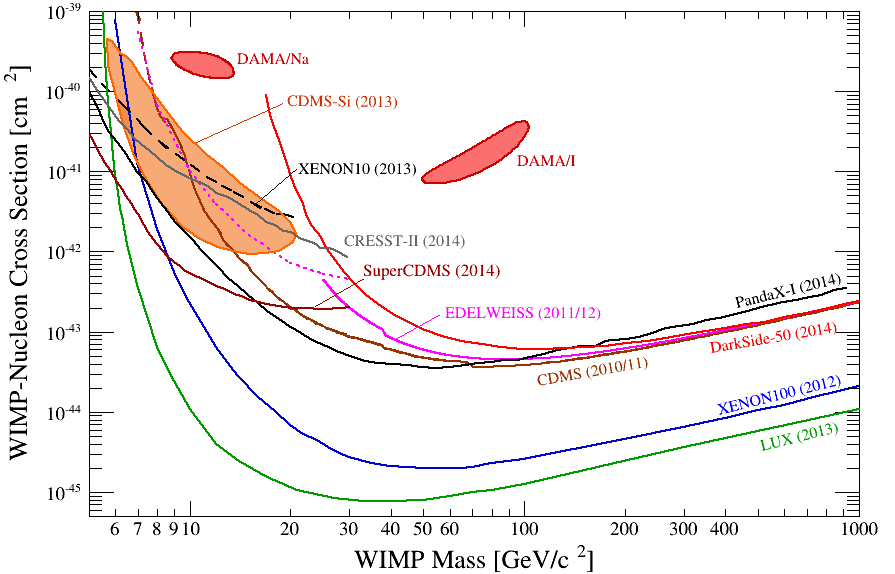}
\caption{Results on spin-independent (scalar) WIMP-nucleon interactions as derived from direct detection experiments. The results from DAMA/LIBRA~\cite{ref::dama,ref::dama_savage} and CDMS-Si~\cite{ref::cdms_si} (closed contours, see text), which could be interpreted as being induced by interactions of light-mass WIMP dark matter, are challenged by the exclusion limits (at 90\% CL, lines) from many other experiments. The parameter space is currently dominated by the dual-phase liquid xenon time projection chambers (TPCs) XENON100~\cite{ref::xe100run10} and LUX~\cite{ref::lux}. Some results discussed in the text are not plotted to increase the readability of the plot. \label{fig::silimits}
}
\end{figure*}

\section{Status direct WIMP Searches}
\label{sec::wimps}

In this Section we present the current status of direct searches for WIMP dark matter. The discussion is separated into spin-independent (scalar) and spin-dependent (axial-vector) interactions. No convincing sign of WIMP dark matter has been observed so far.

\subsection{Spin-independent Interactions}
\label{sec::si}

The current situation concerning spin-independent WIMP-nucleon interactions is summarized in Figure~\ref{fig::silimits}. The most sensitive exclusion limits from 6\,GeV/$c^2$ up to $>$10\,TeV/$c^2$ still come from the liquid xenon experiments XENON100~\cite{ref::xe100run10} and LUX~\cite{ref::lux}. At lower WIMP masses, the best limits are new results from SuperCDMS (3.5--6\,GeV/$c^2$)~\cite{ref::supercdms} and CRESST-II (below 3.5\,GeV/$c^2$)~\cite{ref::cresst}. A few more competitive results from new players in the field have been added as well and will be discussed below, however, the most striking difference compared to the status of 2013~\cite{ref::dm2013} is that the number of ``anomalies'' has been reduced.

Out of the previously reported four anomalies from DAMA/LIBRA~\cite{ref::dama}, CoGeNT~\cite{ref::cogent_2013}, CRESST-II~\cite{ref::cresst_2012} and CDMS-Si (silicon detectors of the CDMS-II experiment)~\cite{ref::cdms_si}, which could all be interpreted as possible hints for the detection of WIMPs with masses around 10\,GeV/$c^2$, only DAMA/LIBRA and CDMS-Si remain.

\paragraph{Low-mass region, scintillators and cryogenic detectors}

The DAMA/LIBRA experiment running at the Gran Sasso National Laboratory (LNGS) in Italy searches for a dark matter-induced annual modulation~\cite{ref::modulation} of the background rate in a massive high-purity NaI(Tl) scintillator target. The collaboration observed such a signal over now 14~annual cycles at 9.3\,$\sigma$ significance, exploiting a cumulative exposure of 1.33\,t\,$\times$\,y~\cite{ref::dama}. The measured phase is is agreement with the expectation of a standard galactic WIMP halo. If this observation is interpreted as being due to WIMP scatters~\cite{ref::dama_savage}, it leads to two preferred regions in the spin-independent parameter space, around 10\,GeV/$c^2$ and 70\,GeV/$c^2$ for interactions with Na or I, respectively. Even though the NaI-crystals are of unprecedented radio-purity, the DAMA/LIBRA background is significantly higher compared to other experiments, and there is no discrimination between ER and NR signals. 

The preferred region derived from data taken with the silicon detectors of the CDMS-II experiment~\cite{ref::cdms_si}, acquired 2007/8 at the Soudan mine (USA), also covers low WIMP masses, however, at considerably lower cross sections compared to DAMA/LIBRA. Three events were observed in this measurement, with an expectation of $\sim$0.5\,background events for the 140\,kg\,$\times$d exposure. A profile likelihood analysis yields only a small probability of 0.2\% for the background-only hypothesis and obtains the highest likelihood for a 8.6~\,GeV/$c^2$ WIMP at $\sigma_n=1.9 \times 10^{-41}$\,cm$^2$. Both hints for a WIMP signal are in conflict with various other results.

One of these is from the same collaboration, using the new SuperCDMS detector. This instrument, based on germanium crystals operated at cryogenic temperatures, features excellent background rejection capabilities by comparing ionization and phonon (heat) signal, which are both simultaneously measured for each event~\cite{ref::scdms_rej}. 11\,events were observed in a 577\,kg\,$\times$\,d exposure, in agreement with the background prediction after taking into account some peculiarities. This leads to an 90\% CL upper limit of $1.2 \times 10^{-42}$\,cm$^2$ at 8\,GeV/$c^2$, almost one order of magnitude below the CDMS-Si best-fit point~\cite{ref::supercdms}.

During the last years, the result from the CoGeNT p-type germanium detector installed at Soudan has received lots of attention, as the observed excess of events could be interpreted as being induced by low-mass WIMPs as well. In such detectors, the pulse rise-time allows the distinction of surface from bulk events to efficiently reduce the background level, however, ER vs.~NR discrimination is not possible. Three new analyses of the data have been presented in~2014, and all lead to the conclusion that the excess is most likely caused by background surface-events.  The first one was published by the CoGeNT collaboration~\cite{ref::cogent_2014}: the dark matter signal extracted from the data in a maximum likelihood analysis remains below the 3\,$\sigma$ level when systematic uncertainties how to model the pulse rise-time are included.  Another analysis from Davis et al.~\cite{ref::davis} concludes that the significance of the excess is even below 1\,$\sigma$, if the a-priori unknown shapes of the rise-time distributions for bulk and surface are systematically varied. Finally, Kelso et al.~\cite{ref::kelso} presented a maximum likelihood analysis in which the best fit to the public CoGeNT data was achieved by a model without a WIMP signal. The fit also gets worse if a dark matter stream is included in the analysis instead of the standard halo model.

The WIMP interpretation of the CoGeNT excess has also been questioned by two other experiments, CDEX and MALBEK, which essentially use the same technology based on p-type, low-threshold, high purity germanium (HPGe) detectors.
The CDEX-1 detector at the Jinping Laboratory (China) is embedded in a NaI(Tl) crystal operated in anti-Compton mode. The background spectrum above 475\,eV observed in 54\,kg\,$\times$\,d agrees with the background model and challenges the CoGeNT excess as being due to dark matter~\cite{ref::cdex}. 
Using their CDEX-0 apparatus, the collaboration published a limit from a run featuring an even lower threshold of $E_{low}=144$\,eV~\cite{ref::cdex_low}. This result is not yet competitive, however, is could explore the region from 1.5--10\,GeV/$c^2$ at cross-sections around $10^{-41\ldots42}$ in the future.
MALBEK is an R\&D project for the MAJORANA experiment, installed at the Kimballton Underground Research Facility (USA). This broad-energy low-background HPGe detector is very similar to CoGeNT and has also been used to set limits on low-mass WIMPs in the parameter space preferred by the previous CoGeNT analysis~\cite{ref::malbek}.

The last anomaly reported in the previous years was from the CRESST-II experiment, measuring with CaWO$_4$ crystals at mK-temperatures at LNGS. These operating conditions allow for the simultaneous measurement of the scintillation light and the heat deposited in the crystal by particle interactions. Comparing the size of the two signals is used to discriminate between ER background and NR signal, while the composite target including light (O), intermediate (Ca) and heavy (W) nuclei provides a good sensitivity from lowest to highest WIMP masses. After the results from 2012~\cite{ref::cresst_2012}, where a significant excess of events above a large background expectation (with a considerable contribution from degraded $\alpha$-events) was observed, the collaboration has recently improved the crystal purity and the experimental design. A new result, based on a single upgraded detector module and a small exposure of 29\,kg\,$\times$\,d, does not confirm the previously reported excess. It also excludes parameter space at low WIMP masses below 3\,GeV/$c^2$, which was previously not covered by direct detection WIMP searches.

\paragraph{Liquid noble gas detectors}

\begin{figure}[b!]
\centering
\includegraphics[width=0.48\textwidth]{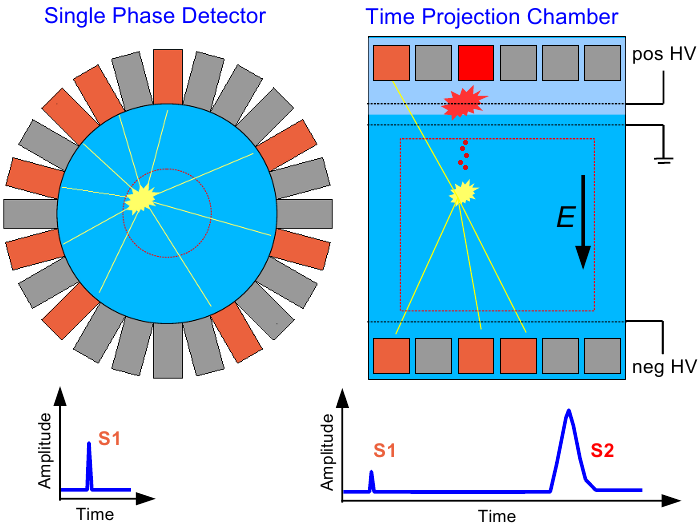}
\caption{Concepts used for dark matter detectors based on liquid noble gases. {\bf (Left)} Single phase detectors filled with liquid xenon, argon or neon record the scintillation light (S1) by means of photodetectors (usually photomultipliers) surrounding the target in $4\pi$. The hitpattern on the sensors allows the reconstruction of the event vertex, which is employed for fiducialization. In detectors filled with liquid argon, the light pulse shape can be used to distinguish ER background from NR signals. {\bf (Right)} Liquid xenon and argon can be easily ionized what is exploited in dual-phase time projection chambers. Particle interactions create prompt scintillation light (S1), detected by two arrays of photosensors above and below the target, and free ionization electrons. These are moved towards the gas phase above the liquid target by an electric field. A second field extracts the electrons into the gas phase, where they generate a secondary scintillation signal (S2),  proportional to the charge signal. The hitpattern on the top array ($xy$) combined with the time difference between S1 and S2 signal ($z$) is used to precisely determine the event position and multiplicity. The ratio S2/S1 is used as a discriminant between ER and NR events.}
\label{fig::detectors}
\end{figure}

The most sensitive dark matter results so far have been published by the XENON100 and LUX collaborations. Both groups operate dual-phase time projection chambers (TPCs)~\cite{ref::dualphase} filled with the the noble gas xenon, which is cooled down to about $-90^\circ$C at 2\,bar pressure such that it liquefies ($\rho \approx 3$\,g/cm$^3$). The principle of such a position-sensitive detector is detailed in Figure~\ref{fig::detectors} (right).

Using liquid xenon as WIMP target combines several advantages~\cite{ref::dualphase_ms}: the high atomic number $A \approx 131$ for an excellent WIMP sensitivity; the high $Z=54$ for dense, compact detectors providing self-shielding against external backgrounds; the absence of any long-lived xenon radio-isotopes (besides $^{136}$Xe which undergoes double-beta decay with $T_{1/2}=2.1 \times 10^{21}$\,y); the high light output at 178\,nm (comparable to NaI(Tl)); and the fact that the target mass can be increased more easily compared to detectors using crystals. Threshold energies of 3\,keVr have already been achieved~\cite{ref::lux} and the combination of external shields, radio-pure detector materials, fiducialization and multiple-scatter rejections leads to demonstrated background rates of $\leq$0.005\,events/keV/kg/day, dominated by decays of the target intrinsic isotopes $^{85}$Kr and $^{222}$Rn. The different energy loss $dE/dx$ for electronic and nuclear recoils leads to different charge-to-light ratios in dual-phase TPCs, which is used to reject $\sim$99.5\% of ERs at $\sim$50\% NR acceptance~\cite{ref::xe100analysis}.

XENON100 is a dual-phase TPC using 62\,kg of liquid xenon as WIMP target~\cite{ref::xe100instr}. Its total xenon mass is 161\,kg, with the xenon outside of the TPC being instrumented as active veto to reject background events. The instrument is running at LNGS (Italy) since 2009 and has reached its design goal by excluding spin-independent WIMP-nucleon cross sections of $2 \times 10^{-45}$\,cm$^2$ at WIMP masses of 50\,GeV/$c^2$ with an exposure of 34\,kg\,$\times$\,225\,d~\cite{ref::xe100run10}. The same low-background dataset has also been used to derive constraints on spin-dependent interactions (see Section~\ref{sec::sd}) and axions (see Section~\ref{sec::other}).

This result has been recently confirmed and superseded by the larger LUX detector located at SURF (USA), which features 250\,kg of liquid xenon inside the TPC (370\,kg total). No excess above the background expectation has been observed in a 118\,kg\,$\times$\,85\,d run, leading to the strongest spin-independent limits which have been published so far for WIMP masses of $>$6\,GeV/$c^2$ ~\cite{ref::lux}.

A new dual-phase TPC has published its first science results in 2014: PandaX-I is installed in the Chinese Jinping Laboratory and has a liquid xenon target of 120\,kg. In contrast to XENON and LUX, which have an aspect ratio close to unity, the PandaX TPC is pancake-shaped in order to optimize the detector for a high light yield, which in turn leads to a low threshold. Consequently, the new limits from a measurement of 37\,kg\,$\times$17\,d go beyond the ones of XENON100 and LUX at very low WIMP masses~\cite{ref::panda_res}. The drawback is a larger background due to the reduced fiducialization power.

First results have also been reported from the DarkSide-50 detector, a dual-phase TPC filled with~46\,kg of liquid argon~\cite{ref::ds-50}. It is installed inside a 30\,t liquid scintillator veto, which itself is located inside a large \u{C}erenkov muon veto. Compared to the very expensive xenon, argon is much cheaper due to its larger atmospheric abundance. The major challenge is the high contamination of $\sim$1\,Bq of $^{39}$Ar per kg of $^{nat}$Ar, which requires ER discrimination efficiencies which are much better than the $10^{-3}$ levels achievable by using the charge-to-light ratio. DarkSide exploits the different scintillation pulse-shape from ERs and NRs~\cite{ref::psd} to efficiently reject ERs. In a 1422\,kg\,$\times$\,d exposure of atmospheric argon, less than 0.1\,event from $^{39}$Ar is expected to leak into the WIMP search region from 38-206\,keVr. This rather high threshold is required as a sizeable amount of photoelectrons (here: 80) needs to be detected in order to be able to trace the pulse shape. This leads to a considerably reduced WIMP sensitivity below $\sim$50\,GeV/$c^2$. No excess of events were observed, the exclusion limit is shown in Figure~\ref{fig::silimits} as well.

\paragraph{Future: Upcoming results and large detectors}

An alternative way to build WIMP detectors using liquefied noble gases is illustrated in Figure~\ref{fig::detectors} (left): single-phase detectors only detect the prompt scintillation light. Compared to TPCs, this has some advantages: the target can be surrounded by light sensors in $4\pi$; the light yield increases due to the absence of quenching effects in the electric field; and the operation of a detector without his bias-voltages is facilitated. The drawback is the lower position resolution and -- in case of a xenon target -- that there are virtually no means to discriminate ERs from NRs. 

XMASS, located in the Kamioka mine (Japan), is an operational single-phase detector, where 835\,kg of liquid xenon are viewed by 642~photomultipliers~\cite{ref::xmass}. First results on WIMPs have already been published~\cite{ref::xmass_wimps}, but were not competitive due to an increased background. This has been improved in the meantime and new results are expected soon. The large single phase project DEAP-3600, with a 3.6\,t liquid argon target, is currently being commissioned at SNOLAB (Canada). First results are expected for 2015, with an ultimate sensitivity around $1 \times 10^{-46}$\,cm$^2$~\cite{ref::deap}.

In July~2014, the US funding agencies DOE and NSF announced a long-awaited decision regarding their joint program for second-generation dark matter detectors~\cite{ref::downselection}. The agencies will support three projects: LZ (LUX-Zeplin)~\cite{ref::lz}, a dual-phase TPC with a 7\,t liquid xenon target, is the successor of the LUX experiment and will be installed in the SURF laboratory (USA) as well. While LZ will mainly search for WIMP dark matter above 10\,GeV/$c^2$, with an optimal sensitivity to spin-independent cross-sections of a few 10$^{-48}$\,cm$^2$, the second experiment, SuperCDMS~\cite{ref::supercdms2}, will focus on the low mass region. It will initially operate $\sim$50\,kg of high purity germanium and silicon crystals at SNOLAB (Canada), and will be designed such that an upgrade to more mass is possible at a later stage. Both experiments should be in the commissioning phase by 2018, and will take several years of data in order to reach their ultimate sensitivity. The third project in the US program is ADMX-Gen2~\cite{ref::admx}. It does not search for WIMP dark matter but operates a microwave cavity in order to look for axions, an alternative dark matter particle which arises in a possible solution to the strong-$CP$ problem~\cite{ref::axion}, see also Section~\ref{sec::other}.

There are more next generation dark matter projects which are currently in the initial design phase, such as DarkSide-G2~\cite{ref::ds-g2}, a dual-phase TPC filled with 3.6\,t of liquid argon, or the upgrades of the PandaX liquid xenon experiment~\cite{ref::pandax_ucla2014}: it first aims at a 500\,kg detector, followed by an even larger stage. 

The next phase of the XENON program is currently being installed underground at LNGS (Italy): the XENON1T detector~\cite{ref::xe1tmv} is a dual-phase TPC with a target mass of 2.0\,t of liquid xenon (dimensions: $\sim$1\,m height and diameter, total mass: 3.3\,t), instrumented by 248~low background photomultipliers~\cite{ref::pmtpaper}. The background goal is $<$1\,event for a 1\,t\,$\times$\,2\,y exposure and will be achieved by careful selection of low-background materials, shielding by a 9.6\,m diameter water shield operated as muon veto as well as by liquid xenon, and by using the charge-to-light ratio for discrimination. Detector commissioning is planned for the second half of 2015, the sensitivity goal of $2 \times 10^{-47}$\,cm$^2$ for $m_\chi \sim 50$\,GeV/$c^2$ can be achieved after 2\,years of operation. All major detector components of XENON1T are designed such that an upgrade to a total xenon mass of $\sim$7\,t is straightforward. This phase, XENONnT, will increase the sensitivity by almost another factor of~10.

\subsection{Spin-dependent Interactions}
\label{sec::sd}

If the WIMP couples to the unpaired nuclear spins of the target nucleus via an axial-vector current, the cross section does not simply scale with $A^2$ as for coherent spin-independent interactions, but depends on a factor $\lambda^2 = J/(J+1) \ ( a_p \langle S_p \rangle + a_n \langle S_n \rangle )^2$, see Eq.~(\ref{eq::sd}). This factor is non-zero only for nuclei with an odd number of protons or neutrons, and is maximal for $^{19}$F ($\lambda^2=0.86$), followed by $^7$Li ($\lambda^2=0.11$), which both have unpaired proton-spins. Some of the experiments described in Section~\ref{sec::si} contain isotopes which are sensitive to spin-dependent interactions, even though to a lesser extent than $^{19}$F. These are $^{23}$Na and $^{127}$I (unpaired protons) as present in DAMA/LIBRA, $^{29}$Si and $^{73}$Ge (unpaired neutrons) in CDMS, and $^{129}$Xe and $^{131}$Xe (unpaired neutrons) in XENON.

\begin{figure}[h!]
 \centering
 \includegraphics[width=0.48\textwidth]{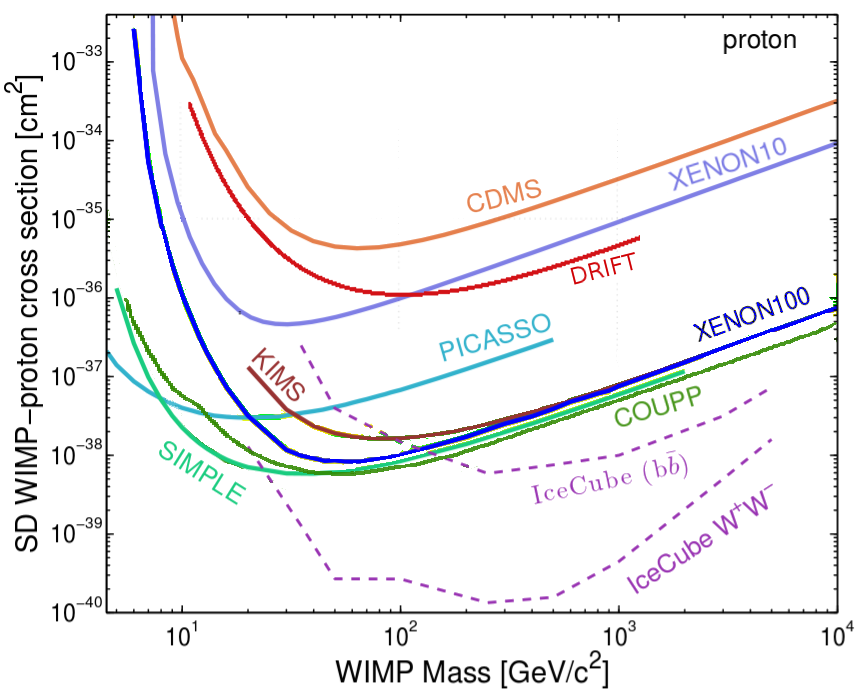}
 \includegraphics[width=0.48\textwidth]{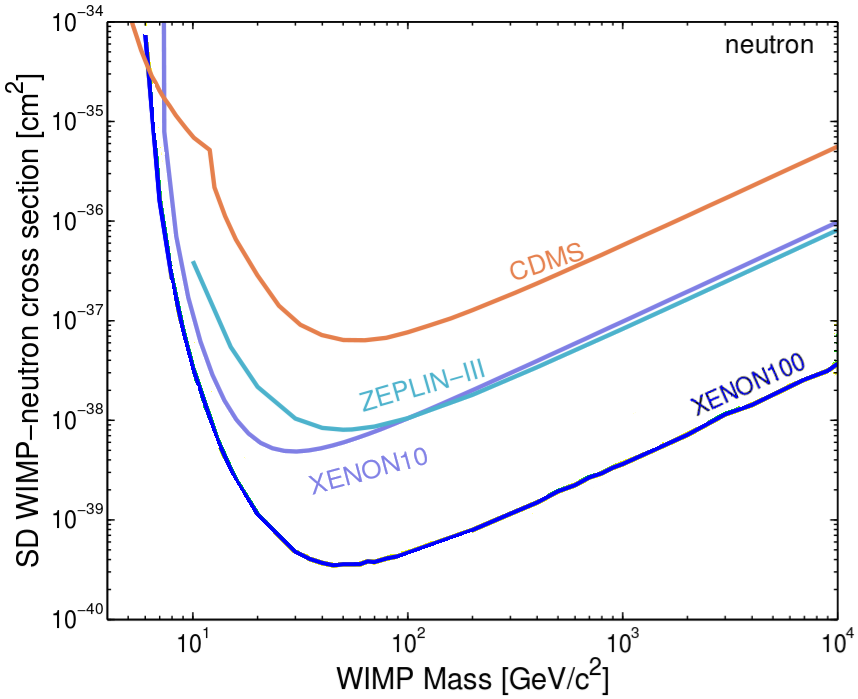}
 \caption{Results on spin-dependent WIMP-nucleon scattering cross sections, presented assuming that WIMPs would couple only to proton- or to neutron-spins. {\bf (Top)} The proton-only case is dominated by results from experiments which employ a target containing $^{19}$F (COUPP~\cite{ref::coupp}, SIMPLE~\cite{ref::simple}, PICASSO~\cite{ref::picasso}). The new results from the directional DRIFT detector~\cite{ref::drift_res} and from indirect WIMP searches by IceCube~\cite{ref::icecube_sd} are also shown. {\bf (Bottom)} The best limit on neutron-only couplings is from XENON100~\cite{ref::xe100sd} using $^{129}$Xe and $^{131}$Xe as target nuclei. Figures~adapted from~\cite{ref::xe100sd}, see more references there.}
 \label{fig::sd}
\end{figure}

The parameter space of spin-dependent WIMP-proton couplings (assuming that $a_n$\,=\,0 in Eq.~(\ref{eq::sd})) is therefore dominated by experiments using targets which contain $^{19}$F, see Figure~\ref{fig::sd} (top). The tightest constraints on the cross section come from COUPP~\cite{ref::coupp}, a bubble chamber filled with CF$_3$I, as well as SIMPLE~\cite{ref::simple} and PICASSO~\cite{ref::picasso}. These consist of superheated droplets of C$_2$CIF$_5$ and C$_4$F$_{10}$, respectively, embedded in a gel. The droplets work as ``mini'' bubble chambers, where incident radiation causes the formation of bubbles, which are detected acoustically and -- in case of COUPP -- also optically. The advantage of this technology is that the detectors can be made almost insensitive to ER background radiation by choosing the right detector parameters (temperature and pressure), while the characteristics of the sound signal can be used to partially discriminate between NRs and $\alpha$-particles~\cite{ref::alphadisc}. In order to keep this forefront position also in the future, PICASSO and COUPP recently merged to form the PICO collaboration, aiming towards a ton-scale bubble chamber, operated with either a CF$_3$I or a C$_3$F$_8$ target.

An interesting new result comes from the DRIFT-IId detector. While all projects discussed so far measure only the energy of an interaction, as well as the particle type and multiplicity in some cases, DRIFT also detects the direction of the recoil in a low-pressure gas TPC. This allows the distinction of WIMP-induced recoils, whose direction is expected to be correlated with the rotation of the Earth, from backgrounds. The detector was filled with a gas mixture of CS$_2$:CF$_4$:O$_2$ at a pressure ratio 30:10:1, searching for spin-dependent WIMP interactions with $^{19}$F. No event was observed in a background-free run observing 33\,g of fluorine gas over 46\,d~\cite{ref::drift_res}.

\begin{figure*}
\centering
\includegraphics[width=0.8\textwidth]{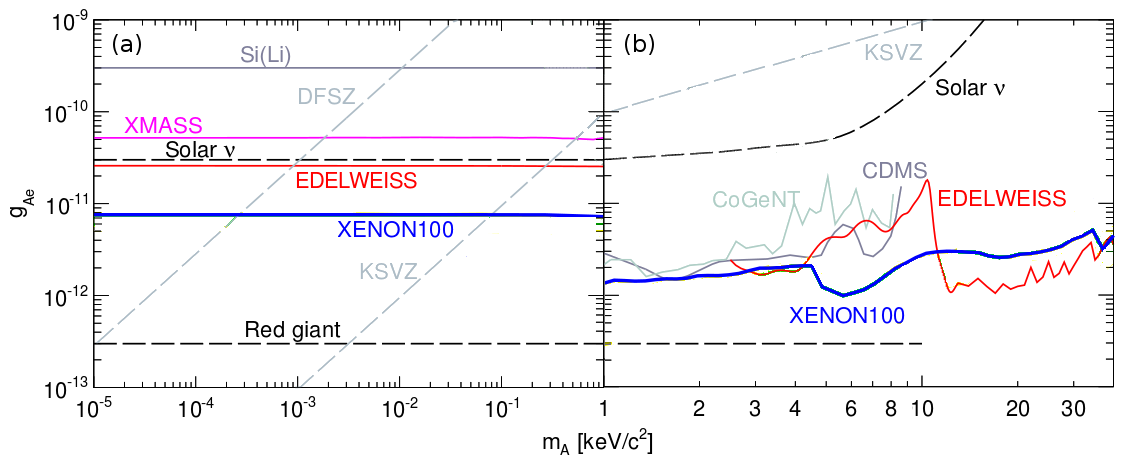}
\caption{Limits on the axio-electric coupling constant $g_{Ae}$ for \textbf{(a)} interactions of solar axions and \textbf{(b)} galactic axion-like particles obtained by dark matter detectors using germanium (EDELWEISS, CoGeNT, CDMS) and liquid xenon targets (XMASS, XENON100). All constraints cut significantly into the dark matter axion regime, as indicated by the DFSZ and KSVZ model lines. Figure adapted from~\cite{ref::xe_axions}, see references to the individual results there. }
\label{fig::axions}
\end{figure*}

About 50\% of the naturally abundant xenon isotopes are the neutron-odd $^{129}$Xe and $^{131}$Xe, which are sensitive targets for spin-dependent neutron-only interactions (assuming $a_p=0$). The most sensitive published exclusion limit is from the XENON100 dual-phase liquid xenon TPC, see Figure~\ref{fig::sd} (bottom), which did not observe a WIMP signal in a 34\,kg\,$\times$\,225\,d exposure~\cite{ref::xe100sd}.

\section{Other Dark Matter Channels}
\label{sec::other}

The discussion so far has focused on weakly interacting massive particles (WIMPs) as candidates for dark matter. Due to their very low radioactive backgrounds, many of the WIMP detectors mentioned so far can also be employed to search for other rare events as well, even neutrino channels will eventually be accessible with multi ton-scale liquid xenon detectors~\cite{ref::neutrinos}. 

Axions are very light dark matter candidates~\cite{ref::axion}, which arise naturally in the Peccei-Quinn solution to the ``strong'' $CP$-problem in QCD, manifesting itself in the absence of an electric dipole moment of the neutron. The most common way to search for axions is via the Primakov effect, where an axion is converted into a photon of the same energy in a magnetic field perpendicular to the axion momentum. Dedicated instruments for axion-searches, such as the microwave cavity ADMX~\cite{ref::admx}, rely entirely on this effect and are installed inside strong magnets. Low-background detectors optimized for WIMP searches do not employ external magnetic fields, hence this effect can only be exploited in crystals with a known orientation axis with respect to a possible axion source, e.g., the Sun, relying on the fields between the nuclei. Limits on $g_{A\gamma}$, the axio-photon coupling constant, have been derived by several germanium experiments (e.g.,~EDELWEISS~\cite{ref::edw_axions}) and NaI(Tl)-crystals (DAMA~\cite{ref::dama_axions}).

Non-crystal targets, as used in the massive liquid xenon detectors, are insensitive to $g_{A\gamma}$ due to the absence of a well-oriented magnetic field. However, these instruments can search for axions which have converted into detectable electrons by the axio-electric effect, hence placing limits on $g_{Ae}$, the coupling constant of axions to electrons. Such as search uses the ER data which is rejected for the WIMP search and requires a very low background even without ER discrimination. Several experiments have recently performed such an analysis, among them the single-phase liquid xenon detector XMASS~\cite{ref::xmass_axions};  the HPGe detectors EDELWEISS~\cite{ref::edw_axions}, CDMS-II and CoGeNT; and the dual-phase liquid xenon TPC XENON100~\cite{ref::xe_axions}. Their limits are shown in Figure~\ref{fig::axions} for two different axion candidates. The first search focuses on solar axions emitted by the Sun, where they are expected to be produced in large amounts. The second one places constraints on galactic axion-like particles (ALPs), assuming that they constitute the entire amount of the observed dark matter. These ALPs do not solve the strong $CP$-problem and are more massive than the classical axion.

A somewhat similar search for bosonic super-WIMPs has been performed by XMASS~\cite{ref::xmass_bosonic}. These warm dark matter candidates with masses of a few~keV/$c^2$ could be absorbed by the xenon atoms, depositing their rest mass energy in the single-phase detector. Due to its very low ER background of $\sim$10$^{-4}$\,events/keV/kg/d, XMASS places tight limits on super-WIMPs with masses between 40 and 120\,keV/$c^2$, as no excess of events was observed in a measurement of 41\,kg\,$\times$\,166\,d. In particular, the possibility that all dark matter is made up from vector super-WIMPs is completely excluded by this result.

\section{Conclusions}

Even though the the best constraints on spin-independent (Figure~\ref{fig::silimits}) and spin-dependent (Figure~\ref{fig::sd}) WIMP-nucleon interactions were not improved in~2014, it has been a very interesting year in terms of direct searches for WIMP dark matter, as several new detectors came online or presented first science results. Especially interesting is the apparent ``clean-up'' of the low-mass WIMP region, where some of the previous anomalies have disappeared (CRESST) or considerably lost statistical significance (CoGeNT). As most of the WIMP detectors discussed above are still operational and continue to take data, improved results are expected in the upcoming years, for WIMP and non-WIMP dark matter channels.

At the same time, competitive new detectors approaching ton-scale target masses are under commissioning (DEAP-3600), under construction (XENON1T), or in the design phase (LZ, PICO, DarkSide-G2, etc.). These projects will significantly improve the sensitivity to WIMP-nucleon interactions by 1-2~orders of magnitude compared to the present status, shedding light at one of the most important topics in astroparticle physics.


\end{document}